\documentstyle[prl,aps,epsf,floats]{revtex}

\begin{document}
\draft

\twocolumn[\hsize\textwidth\columnwidth\hsize\csname @twocolumnfalse\endcsname

\title{Spin extraction from a nonmagnetic semiconductor: Tunneling of 
electrons from semiconductors into ferromagnets through a modified Schottky barrier}
\author{A.M. Bratkovsky and V.V. Osipov}
\address{
Hewlett-Packard Laboratories, 1501 Page Mill Road, 1L, Palo Alto, CA 94304
}
\date{October 8, 2003 }
\maketitle

\begin{abstract}

New efficient mechanism of obtaining spin polarization in 
{\em nonmagnetic} semiconductors at arbitrary temperutures is described.
The effect appears during tunneling 
of electrons from a nonmagnetic semiconductors (S) into ferromagnet (FM) 
through a Schottky barrier modified with very thin heavily doped interfacial layer. 
We show that electrons with a certain spin projection are extracted from S, 
while electrons with  the opposite spins are accumulated in S. 
The spin density increases and spin penetration  depth decreases with current.  
	
\end{abstract}

\vskip2pc]

\narrowtext

Spintronics, i.e. the manipulation of spin in solid state devices, opens up
the possibilities for designing ultrafast scaleable devices 
\cite{Wolf,Datta}. 
Giant and tunnel magnetoresistance effects in magnetic layered systems
proved to be practically important phenomena \cite{GMR,Slon,Brat}.\ An
injection of spin-polarized carriers into semiconductors is of particular
interest because of relatively large spin-coherence lifetime of electrons in
semiconductors \cite{Aw} and possibilities for applications in ultrafast
devices and quantum computers \cite{Wolf,Datta}. An efficient spin injection
in heterostructures with magnetic semiconductor as a spin source have been
reported in Refs. \cite{MSemi}. However, the Curie temperature of the known
magnetic semiconductors is substantially below room temperature. In Ref. 
\cite{Zut} it was shown that spin injection and extraction in p-n junctions
containing magnetic semiconductors can occur in large magnetic fields (at
least a few Tesla). Fairly efficient spin injection from
ferromagnets (FM) into nonmagnetic semiconductors (NS) has been demonstrated
recently at low temperatures \cite{Jonk}. The attempts to achieve an
efficient room-temperature spin injection from FM into NS have faced
substantial difficulties \cite{Ferro}. Optimal conditions of the spin
injection\ from FM into NS have been discussed in Refs. \cite{Rash}. Spin
diffusion and drift in an electric field have been investigated in Refs. 
\cite{Flat}. The spin polarization of photoexcited electrons in NS due to
reflection off a ferromagnet was studied in Refs. \cite{Ciuti}.

In this paper we consider a ferromagnet-semiconductor junction with very
thin heavily doped semiconductor layer ($\delta -$doped layer) and show that
tunnelling of electrons from a {\it nonmagnetic }semiconductor into a
ferromagnet through the $\delta -$doped layer results in formation of highly
spin-polarized electrons in the semiconductor near the interface at room
temperature in absence of an external magnetic field. We assume that the $%
\delta -$doped layer has the thickness $l\lesssim 2$ nm, and the
concentration of donors, $N_{d}^{+}$, satisfying the condition: $%
N_{d}^{+}l^{2}=2\epsilon \epsilon _{0}(\Delta -\Delta _{0})/q^{2},$ where $q$
is the elementary charge, $\epsilon $ ($\epsilon _{0})$ is the permittivity
of the semiconductor (vacuum), $\Delta _{0}=E_{c}-F$, $F$ the Fermi level, $%
E_{c\text{ }}$the bottom of a semiconductor conduction band, $\Delta $ the
height of potential barrier at the ferromagnetic-semiconductor interface.
(Note that for GaAs and Si $\Delta \simeq 0.5-0.8$ eV practically for all
metals including Fe, Ni, and Co{\it \ }\cite{sze,Jonk}). The energy band
diagram of such a FM$-n^{+}-n-$S structure includes a $\delta -$spike of
height $(\Delta -\Delta _{0})$ and thickness $l$ (Fig. 1).\ We consider that
due to a smallness of $l$ the electrons can easily tunnel through the $%
\delta -$spike.
\begin{figure}[t]
\epsfxsize=3.in \epsffile{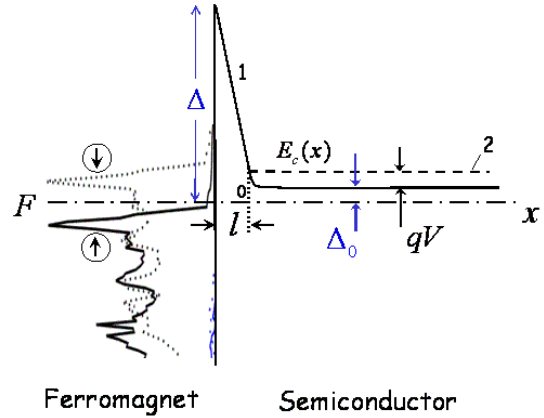}
\caption{Schematic~energy~diagram of ferromagnet-semiconductor
heterostructure with $\delta -$doped layer in equilibrium (curve 1) and at
bias voltage $V$ (curve 2). Here $F$ is the Fermi level, $\Delta $ the height, 
$l$ the thickness of the interface potential barrier due to the $\delta -
$doped layer, $\Delta _{0}$ the height of a barrier in the $n-$type
semiconductor, $E_{c}(x)$ the bottom of conduction band in the
semiconductor. The density of states of spin polarized electrons in
ferromagnetic Ni is shown at $x<0,$ as an example.
}
\label{fig:fig1}
\end{figure}

We assume that the electron energy $E$ and component $\vec{k}_{\parallel }$
of the wave vector parallel to the interface are conserved during tunneling
through the FM-S interface. The current density of electrons with spin $%
\sigma $ from the semiconductor into the ferromagnet at the interface
($x=0$, Fig.~1), which we denote $J_{\sigma 0}=J_{\sigma }(0)$, 
can be written as \cite{Duke,Brat} 
\begin{equation}
J_{\sigma 0}=\frac{q}{h}\int dE[f(E-F_{\sigma 0})-f(E-F)]\int \frac{%
d^{2}k_{\parallel }}{(2\pi )^{2}}T_{\sigma },  \label{cur}
\end{equation}
where $f(E)=[\exp (E-F)/T+1]^{-1}$ the Fermi function, $T$ the temperature
(in units $k_{B}=1$), $T_{k\sigma }$ the transmission probability, the
integration includes a summation with respect to a band index, and we
assume that the negative bias voltage $V$ is applied to the
semiconductor. We need to 
consider, in variance with Refs. \cite{Duke,Brat}, that electrons with spin $%
\sigma $ in the semiconductor can be out of equilibrium with their
distribution described by a Fermi function with a quasi-Fermi level $%
F_{\sigma }(x)$. For definiteness, we consider a nondegenerate semiconductor 
\cite{DS} where a total electron density $n$ and a density of electrons with
spin $\sigma $ near the interface, $n_{\sigma }(0)$, are given by 
\begin{equation}
n=N_{c}\exp \left( -\frac{\Delta _{0}}{T}\right) ,~ n_{\sigma }(0)=%
\frac{N_{c}}{2}\exp \left( \frac{F_{\sigma 0}-E_{c}}{T}\right) .  \label{nc}
\end{equation}
Here $F_{\sigma 0}=F_{\sigma }(0)$, $N_{c}=2M_{c}(2\pi m_{\ast
}T)^{3/2}h^{-3}$ is the effective density of states of the semiconductor
conductor band \cite{sze}, $m_{\ast }$ the effective mass of electrons in
the semiconductor, $M_{c}$ the number of band minima. The analytical
expressions for the transmission probability $T_{\sigma }(E,k_{\parallel })$%
\ can be obtained in an effective mass approximation $\hbar k_{\sigma
}=m_{\sigma }v_{\sigma }$ where $v_{\sigma }$ and $m_{\sigma }$ are the
velocity and the mass of electron with spin $\sigma $. The potential barrier
(Fig. 1) has a ``pedestal'' with a height $(\Delta _{0}+qV).$ For electron
energies $E\gtrsim E_{c}=F+\Delta _{0}+qV$ one can approximate the $\delta -$%
barrier by a triangular shape and find approximately 
\begin{equation}
T_{\sigma }=\frac{16\alpha m_{\sigma }m_{\ast }k_{\sigma x}k_{x}}{m_{\ast
}^{2}k_{\sigma x}^{2}+m_{\sigma }^{2}\kappa ^{2}}e^{-\eta \kappa l}=\frac{%
16\alpha v_{\sigma x}v_{x}}{v_{\sigma x}^{2}+v_{tx}^{2}}e^{-\eta \kappa l},
\label{Ts}
\end{equation}
where $\kappa =(2m_{\ast }/\hbar ^{2})^{1/2}(\Delta +F-E_{x})^{3/2}(\Delta
-\Delta _{0}-qV)^{-1}$, $E_{x}=E-E_{\parallel }$, $E_{\parallel }=\hbar
^{2}k_{\parallel }^{2}/2m_{\ast }$, $v_{t}=\hbar \kappa /m_{\ast }$ is the
``tunneling'' velocity, $v_{x}$ ($v_{\sigma x})$ is the $x-$component of the
velocity of electrons $v$ ($v_{\sigma })$ in the semiconductor
(ferromagnet), $\alpha =\pi (\kappa l)^{1/3}\left[ 3^{1/3}\Gamma ^{2}\left( 
\frac{2}{3}\right) \right] ^{-1}\simeq 1.2(\kappa l)^{1/3},$ $\eta =4/3$
(cf. $\alpha =1$ and $\eta =2$ for a rectangular barrier). The
preexponential factor in Eq.\ (\ref{Ts}) accounts for a mismatch of the
effective masses, $m_{\sigma }$ and $m_{\ast }$, and the velocities, $%
v_{\sigma x}$ and $v_{x}$, of electrons at the FM-S interface. Note that Eq.
(\ref{Ts}) is similar to Eq.~(2) of Ref. \cite{Brat}.

In a regime of interest, $T<\Delta _{0}\ll \Delta $ and $E\gtrsim
E_{c}=(F+\Delta _{0}+qV)>F$ \cite{DS}, Eqs. (\ref{cur}) and (\ref{Ts}) can
be written, accounting for a singular energy dependence of the velocity in
the nondegenerate semiconductor $v_{x}=\sqrt{2(E_{x}-E_{c})/m_{\ast }},$ as\ 
\begin{eqnarray}
J_{\sigma 0} &=&\frac{2^{3/2}\alpha qM_{c}}{\pi ^{3/2}\hbar m_{\ast }^{1/2}}%
\left( e^{\frac{F_{\sigma 0}}{T}}-e^{\frac{F}{T}}\right) \int_{0}^{\infty
}dk_{\parallel }^{2}\int_{E_{c}+E_{\parallel }}^{\infty }dE  \nonumber \\
&&\times \frac{v_{\sigma x}}{v_{\sigma x}^{2}+v_{tx}^{2}}(E-E_{c}-E_{%
\parallel })^{1/2}\exp \left( -\eta \kappa l-\frac{E}{T}\right)   \nonumber
\\
&=&\frac{2^{3/2}\alpha qM_{c}m_{\ast }^{1/2}v_{\sigma 0}T^{5/2}}{\pi
^{3/2}\hbar ^{3}(v_{t0}^{2}+v_{\sigma 0}^{2})}e^{-\eta \kappa _{0}l-\frac{%
E_{c}}{T}}\left( e^{\frac{F_{\sigma 0}}{T}}-e^{\frac{F}{T}}\right) ,
\label{J}
\end{eqnarray}
where $\kappa _{0}\equiv 1/l_{0}=(2m_{\ast }/\hbar ^{2})^{1/2}(\Delta
-\Delta _{0}-qV)^{1/2}$, $v_{t0}=\sqrt{2(\Delta -\Delta _{0}-qV)/m_{\ast }}$
and $v_{\sigma 0}=v_{\sigma }(\Delta _{0}+qV)$. From Eqs. (\ref{nc}), and (%
\ref{J}) we find 
\begin{equation}
J_{\sigma 0}=jd_{\sigma }\left( \frac{2n_{\sigma }(0)}{n}-e^{-\frac{qV}{T}%
}\right) ,\text{ }j=\frac{4\alpha _{0}}{3}nqv_{T}e^{-\eta \kappa _{0}l}.
\label{J0}
\end{equation}
We have introduced the thermal velocity $v_{T}\equiv \sqrt{3T/m_{\ast }}$, $%
\alpha _{0}=1.2(\kappa _{0}l_{+})^{1/3},$ and the spin factor $d_{\sigma
}=v_{T}v_{\sigma 0}(v_{t0}^{2}+v_{\sigma 0}^{2})^{-1}$. One can see
from Eq. (\ref{J0}) that the current of electrons flowing from nonmagnetic
semiconductors into ferromagnets $J_{\sigma 0}$\ depends on an electron spin 
$\sigma .$ Note that the present expression for the current is in stark
difference from standard expressions for a current through Schottky
metal-semiconductor contact (see Ref. \cite{sze}).

The density of electrons with spin $\sigma $ and their spatial distribution
in the semiconductor near the interface is determined by the continuity
equation \cite{sze,Flat} 
\begin{equation}
dJ_{\sigma }/dx=q\delta n_{\sigma }/\tau _{s},  \label{CC}
\end{equation}
where $\delta n_{\sigma }=n_{\sigma }-n/2$ and $\tau _{s}$ the electron
spin-coherence time. The current density\ of electrons is given by the usual
expression

\begin{equation}
J_{\sigma }=qD(dn_{\sigma }/dx)+q\mu n_{\sigma }E,  \label{Cur}
\end{equation}
where $D$ ($\mu )$ are the diffusion constant (mobility) of the electrons, $E
$ is electric field. Both $J$ and $E$ are directed\ along the $x-$axis, and $%
\tau _{s}$ $\gg \tau _{p}$ , i.e. $L_{s}=\sqrt{D\tau _{s}}\gg \lambda =\sqrt{%
D\tau _{p}}$, $\tau _{p}$ the relaxation time of electron momentum in the
semiconductor. From continuity of the\ total current, $J(x)=J_{\uparrow
}+J_{\downarrow }={\rm const}$, and $n(x)=n_{\uparrow }(x)+n_{\downarrow
}(x)={\rm const}$ we have $E(x)=J/q\mu n={\rm const}$ and $\delta
n_{\uparrow }(x)=-\delta n_{\downarrow }(x)$. Using (\ref{CC}) and (\ref{Cur}%
), we obtain 
\begin{equation}
L_{s}^{2}\frac{d^{2}\delta n_{\uparrow }}{d^{2}x}+L_{E}\frac{d\delta
n_{\uparrow }}{dx}-\delta n_{\uparrow }=0,  \label{Cons}
\end{equation}
where $L_{s}=\sqrt{D\tau _{s}}$ and $L_{E}=$ $\mu \tau _{s}E=\tau _{s}J/qn$
are the spin-diffusion and the drift lengths of electrons in a
semiconductor, respectively \cite{Flat}. The solution of Eq. (\ref{Cons}),
satisfying a boundary condition $\delta n_{\uparrow }\rightarrow 0$
at $x\rightarrow \infty ,$ is 
\begin{equation}
\delta n_{\uparrow }(x)=\frac{n}{2}ce^{-x/L},\text{ }L=\frac{1}{2}\left( 
\sqrt{4L_{s}^{2}+L_{E}^{2}}-L_{E}\right) .  \label{nL}
\end{equation}
The parameter $c$ in the above expression is found as follows.
We obtain from Eqs.~(\ref{J0}), (\ref{nL})
\begin{equation}
J_{\uparrow 0}=j_{0}d_{\uparrow }(\gamma +c)=\frac{J}{2}\frac{\left(
1+P\right) \left( \gamma +c\right) }{\gamma +cP},  \label{J2}
\end{equation}
where $\gamma =1-\exp (-qV/T)$ and we have introduced the effective spin
polarization 
\begin{equation}
P=\frac{d_{\uparrow }-d_{\downarrow }}{d_{\uparrow }+d_{\downarrow }}=\frac{%
(v_{\uparrow 0}-v_{\downarrow 0})(v_{t0}^{2}-v_{\uparrow 0}v_{\downarrow 0})%
}{(v_{\uparrow 0}+v_{\downarrow 0})(v_{t0}^{2}+v_{\uparrow 0}v_{\downarrow
0})},  \label{pol}
\end{equation}
which is the spin polarization of current in a tunneling FM-I-FM structure 
\cite{Brat}. On the other hand, substituting Eq.~(\ref{nL}) into Eqs. (\ref
{Cur}) and using that $J=q\mu En,$ we find
\begin{equation}
J_{\uparrow 0}=\frac{J}{2}\left[ 1+c\left( 1-\frac{D}{\mu EL}\right) \right]
=\frac{J}{2}\left( 1-c\frac{L}{L_{E}}\right) .  \label{J01}
\end{equation}
One obtains a quadratic equation for $c$
from Eqs.~(\ref{J2}) and (\ref{J01}), which has a 
unique physical solution, quite accurately represented as 
\begin{equation}
c=-P\frac{L_{E}(1-e^{-qV/T})}{L_{E}+L(1-e^{-qV/T})}.  \label{C}
\end{equation}

One can see from (\ref{nL}) and (\ref{C}) that at very small bias voltage, $%
qV\ll T$, and current $J\ll J_{s}=qnL_{s}/\tau _{s},$ when $L_{E}=\mu \tau
_{s}E=J\tau _{s}/en\ll L_{s},$ the induced spin polarization in the
semiconductor is small, $c/P\approx -\left( \xi J/J_{s}\right) /(1+\xi ),$
where $\xi =L_{s}/v_{T}\tau _{s}\ll 1$. At large voltages $qV\gg T$ and the
current $J$ approaching its saturation value $J_{max}=J_{s}/\xi \gg J_{s},$ the
polarization of electrons reaches the limiting value $c/P=-2/\left( 1+\sqrt{%
1+4\xi ^{2}}\right) \approx -1,$ and{\bf \ }near the interface $\delta
n_{\uparrow (\downarrow )}(0)\approx \mp Pn/2.$ The spin penetration length
is equal to $L=L_{S}^{2}/L_{E}=enD/J$. Thus, at $P<0$ $(d_{\uparrow
}<d_{\downarrow })$ the electrons with spin $\sigma =\uparrow $ are
accumulated, $n_{\uparrow }(0)\approx (1+\left| P\right| )n/2,$ while\ the
electrons with spin $\sigma =\downarrow $ are extracted, $n_{\uparrow
}(0)\approx (1-\left| P\right| )n/2,$ from a semiconductor. The penetration
length of the induced spin polarization area decreases as $L\propto 1/J$
(Fig. 2). For a typical semiconductor parameters at room temperature $%
D_{n}\approx 25$ cm$^{2}/s$, $\tau _{s}=4\times 10^{-10}$ s, and $L_{s}=1\mu 
$m. Thus, when $J\approx 3J_{s}$ the spin accumulation/extraction at the
interface is $\delta n_{\uparrow (\downarrow )}(0)\approx \mp 0.91Pn/2${\bf %
\ } and $L=0.3\mu $m.
\begin{figure}[t]
\epsfxsize=3.6in \epsffile{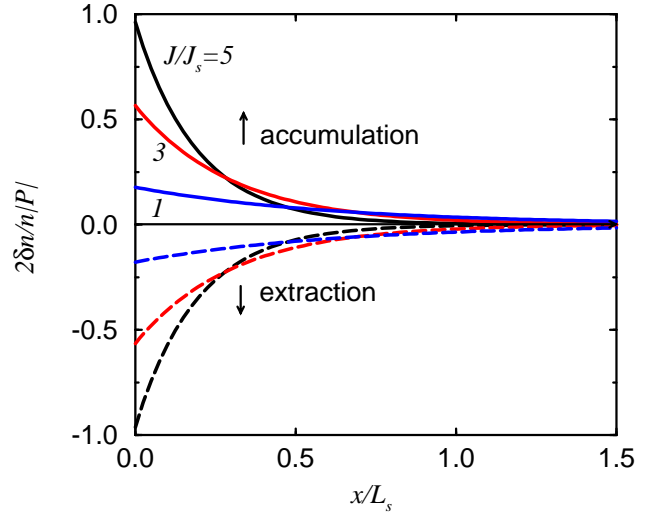}
\caption{Spatial distribution of spin polarized electrons 
$2\delta n_{\uparrow (\downarrow)} (0)/n|P|$ in the
semiconductor at different currents densities $j=J/J_{s}$ shown next
to the lines. 
$J_{s}=qnL_{s}/\tau _{s},$ at $\xi =L_{s}/v_{T}\tau _{s}=0.2$
in a structure like the one shown in Fig.~1.
 }
\label{fig:fig2}
\end{figure}

The required spin-relaxation time is obtained from Eq. (\ref{J0}), which
yields $J\gg J_{s}$ at $qV\gtrsim T$ when the $\delta $-doped layer is very
thin, $l\lesssim l_{0}.$ The corresponding condition reads 
\begin{equation}
\tau _{s}\gg D\left( \frac{\Delta -\Delta _{0}}{4\alpha _{0}v_{\sigma 0}^{2}T%
}\right) ^{2}\exp \frac{2\eta l}{l_{0}}.  \label{Cond2}
\end{equation}
With typical semiconductor parameters at $T\simeq 300$ K ($D_{n}\approx 25$
cm$^{2}$/s, $(\Delta -\Delta _{0})\simeq 0.5$ eV, $v_{\sigma 0}\simeq 10^{8}$
cm/s \cite{sze}) it is satisfied at $l\lesssim l_{0}$ when the
spin-coherence time $\tau _{s}\gg 10^{-12}$ s. We notice that $\tau _{s}$
can be $\sim 1$ ns even at $T\simeq 300$ K (e.g. in ZnSe \cite{Aw}).

We emphasize that $l_{0}\propto (\Delta -\Delta _{0}-qV)^{1/2}$, $v_{\sigma
0}=v_{\sigma }(F+\Delta _{0}+qV),$ and $P$ are all functions of the bias
voltage $V$ and $\Delta _{0}$. Therefore, by adjusting $V$ and $\Delta _{0}$
one may be able to maximize a spin accumulation. This can be achieved by
means of electron tunneling through the $\delta $-doped layer, when the
bottom of the conduction band in a semiconductor $E_{c}=F+\Delta _{0}+qV$ is
close to a peak in the density of states of minority electrons in the
elemental ferromagnet like e.g. Ni, $F+\Delta _{\downarrow }$, $\Delta
_{\downarrow }\simeq 0.1$ eV \cite{Mor}, as illustrated in Fig. 1.


\begin{references}
\bibitem{DS}  Spin extraction from {\em degenerate} highly doped
semiconductors can be achieved at arbitrary temperatures from $T=0$ to $%
T \gtrsim 300$K. In this case $E_{c}$ should be substituted for $F+qV$ in all
equations.
\end{references}
\end{document}